\documentclass{article}
\usepackage {amsmath,amsfonts,amssymb,bbold}
\usepackage {graphics}
\usepackage {graphicx}
\usepackage {longtable}
\usepackage{natbib}
\usepackage {subfigure}
\usepackage {amssymb} 
\usepackage{bm,url}
\newcommand{\ie}{{\em i.e. }}
\newcommand{\eg}{{\em e.g. }}

\title
{\bf Approximate Bayesian Computation scheme for parameter inference and model selection in dynamical systems}

\author
{{\bf Tina Toni$^{1,2}$, David Welch$^3$, Natalja Strelkowa$^4$, Andreas Ipsen$^{5}$}\\ {\bf \& Michael P.H. Stumpf$^{1,2}$} \\
$^{1}$Centre for Bioinformatics, Division of Molecular Biosciences, Imperial College London, UK\\
$^{2}$Institute of Mathematical Sciences, Imperial College London, UK\\
$^{3}$Department of Epidemiology and Public Health, Imperial College London, UK\footnote{Currently at Department of Statistics, Pennsylvania State University, USA.}\\
$^{4}$Department of Bioengineering, Imperial College London, UK\\
$^{5}$Department of Biomolecular Medicine, Imperial College London, UK\\
ttoni@imperial.ac.uk, m.stumpf@imperial.ac.uk}

\begin{document}

\date{} 
\maketitle

\begin{abstract}

Approximate Bayesian computation methods can be used to evaluate posterior distributions without having to calculate  likelihoods. In this paper we discuss and apply an approximate Bayesian computation (ABC) method based on sequential Monte Carlo (SMC) to estimate parameters of
dynamical models.  We show that ABC SMC gives information about the inferability of parameters and model sensitivity to changes in parameters, and tends to perform better than other ABC approaches.  The algorithm is applied to several well known biological systems,
for which parameters and their credible intervals are inferred. Moreover, we develop ABC SMC as a tool for model selection; given a range of different mathematical descriptions, ABC SMC is able to choose the best model using the standard Bayesian model selection apparatus.
\end{abstract}





\section{Introduction}

Most dynamical system studied in the physical, life and social sciences and engineering are modelled by ordinary, delay or stochastic differential equations. However, for the vast majority of systems and particularly for biological systems, we lack reliable
information about parameters and frequently have several competing models for the structure of the underlying equations. Moreover, the biological experimental data is often scarce and incomplete, and the likelihood surfaces of large models are complex. The analysis of such 
dynamical systems therefore requires new, more realistic quantitative and predictive models. Here we will develop novel statistical tools that allow us to analyze such data in terms of dynamical models by (i) providing estimates for model parameter values, and (ii) allowing us to compare the performance of different models in describing the overall data.

In the last decade extensive research has been conducted on estimating the parameters of deterministic systems. Much attention has been given to local and global
nonlinear optimization methods \citep{Mendes:1998p646, Moles:2003p8} and generally parameter estimation has been performed by maximum likelihood estimation \citep{Timmer:2004p474,Muller:2004p265,Baker:2005p678,Bortz:2006p3413}. The methods developed for ordinary differential equations have been extended  to ordinary
differential equations with time delays \citep{Horbelt:2002p475}. Deterministic models have also been parameterized in a Bayesian framework using Bayesian
hierarchical models \citep{Putter:2002p10963,Banks:2005p15510,Huang:2006p11202}. Simulated annealing, which attempts to avoid getting trapped
in local minima, is another well known optimization algorithm that has been found successful in various applications \citep{Kirkpatrick:1983p611,Mendes:1998p646}. There are also  several Monte Carlo based
approaches applied to parameter estimation of deterministic \citep{Battogtokh:2002p544, Brown:2003p727} and stochastic \citep{Sisson:2007p2} systems. Parameter estimation for
stochastic models has been extensively explored in financial mathematics \citep{Johannes:2005p7448} and has been applied to biological
systems in a frequentist maximum likelihood \citep{Reinker:2006p374} and Bayesian \citep{Golightly:2005p372,Golightly:2006p11518,Wilkinson:2006p2403} framework. 

Most commonly, model selection has been performed by likelihood ratio tests (in case of nested models) or the Akaike information criterion (in case of non-nested models). Recently Bayesian methods have increasingly been  coming into use. \cite{Vyshemirsky:2008p14865} have investigated different ways of computing Bayes factors for model selection of deterministic differential equation models, and  \cite{Brown:2003p727} have used the Bayesian information criterion. In population genetics, model selection has been performed using approximate Bayesian computation (ABC) in its basic rejection form \citep{Zucknick:2004p17144, Wilkinson:2007p14173} and coupled with multinomial logistic regression \citep{Beaumont:2008p17134, Fagundes:2007p1947}. 

There is thus a wide variety of tools available for parameter estimation and, to a lesser extent, model selection. However, to our knowledge, no method available can be applied to all different kinds of modelling approaches (\eg ordinary or stochastic differential equations with and without time delay) without substantial modification, estimate credible intervals, take incomplete or partially observed data as input, be employed for
model selection, and reliably explore the whole parameter space without getting trapped in local extrema. 

In this paper we apply an ABC method based on sequential Monte Carlo (SMC) to the parameter estimation and model selection problem for dynamical models. In ABC methods the evaluation of the likelihood is replaced by a simulation-based procedure \citep{Pritchard:1999p17234,Beaumont:2002p13862,  Marjoram:2003p5, Sisson:2007p2}. We explore the information that ABC SMC gives about inferability of parameters and sensitivity of the model to parameter variation. Furthermore we compare the performance of ABC SMC to other approximate Bayesian computation methods. The method is illustrated on two simulated datasets (one from ecology and one from molecular systems biology), and real and simulated epidemiological datasets. As we will show, ABC SMC yields reliable parameter estimates with credible intervals; can be applied to different types of models (\eg deterministic as well as stochastic models); is relatively computationally efficient (and easily parallelized); allows for discrimination among sets of candidate models in a formal Bayesian model-selection sense; and gives us an assessment of model and parameter sensitivity.

\section{Methods}
In this section we review and develop the theory underlying ABC with emphasis on applications to dynamical systems, before introducing a formal Bayesian model selection approach in an ABC context.

\subsection{Approximate Bayesian Computation}

Approximate Bayesian Computation  methods have been conceived with the aim of inferring posterior distributions where likelihood functions are computationally intractable or too costly to
evaluate. They exploit the computational efficiency of modern simulation techniques by replacing calculation of the likelihood with a comparison between the observed data and simulated data. 

Let $\theta$ be a parameter vector to be estimated. Given the prior distribution $\pi(\theta)$, the goal is to approximate the posterior
distribution, $\pi(\theta|x) \propto f(x|\theta)\pi(\theta)$, where $f(x|\theta)$ is the likelihood of $\theta$ given the data $x$. ABC methods have the following generic form:

\begin{description}

 	\item [1] Sample a candidate parameter vector $\theta^*$ from some proposal distribution $\pi(\theta)$.

	\item [2] Simulate a dataset $x^*$ from the model described by a conditional probability distribution $f(x|\theta^*)$.

	\item [3] Compare the simulated dataset, $x^*$, with the experimental data, $x_0$, using a distance function, $d$, and tolerance $\epsilon$; if $d(x_0, x^*)
	\leq \epsilon$, accept $\theta^*$. The tolerance $\epsilon \geq 0$ is the desired level of agreement between  $x_0$ and $x^*$.
\end{description}
The output of an ABC algorithm is a sample of parameters from a distribution $\pi(\theta|d(x_0, x^*) \leq \epsilon)$. If $\epsilon$ is sufficiently small then the distribution $\pi(\theta|d(x_0, x^*) \leq \epsilon)$ will be a good approximation for the posterior
distribution, $\pi(\theta|x_0)$. It is often difficult to define a suitable distance function $d(x_0,x^*)$ between the full datasets, so one may instead replace it with a distance defined on summary statistics, $S(x_0)$ and $S(x^*)$, of the datasets.  That is, the distance function may be defined as $d(x_0,x^*)  = d'(S(x_0),S(x^*))$, where $d'$ is a distance function define on the summary statistic space. However, as we here consider values of a dynamical process at a set of time points, we are able to compare the datasets directly without the use of summary statistics. In any case, the algorithms take the same form. 
\par
The simplest ABC algorithm is the \textit{ABC rejection sampler} \citep{Pritchard:1999p17234}:
\begin{description}

 	\item [R1] Sample $\theta^*$ from $\pi(\theta)$.

	\item [R2] Simulate a dataset $x^*$ from $f(x|\theta^*)$.

	\item [R3] If $d(x_0, x^*) \leq \epsilon$, accept $\theta^*$, otherwise reject.

	\item [R4] Return to \textbf{R1}.
\end{description}
The disadvantage of the ABC rejection sampler is that the acceptance rate is low when the prior distribution is very different from the posterior distribution. To avoid this problem an ABC method based on Markov chain Monte Carlo was introduced \citep{Marjoram:2003p5}. The \textit{ABC MCMC algorithm} proceeds as follows:

\begin{description}

 	\item [M1] Initialize $\theta_i$, $i=0$.

	\item [M2] Propose $\theta^*$ according to a proposal distribution $q(\theta|\theta_i)$.

	\item [M3] Simulate a dataset $x^*$ from $f(x|\theta^*)$. 
	
	\item [M4] If $d(x_0, x^*) \leq \epsilon$, go to \textbf{M5}, otherwise set $\theta_{i+1} = \theta_{i}$ and go to \textbf{M6}.

	\item [M5] Set $\theta_{i+1} = \theta^*$ with probability
$$
\alpha = \textrm{min} \left( 1, \frac{\pi(\theta^*)q(\theta_i|\theta^*)}{\pi(\theta_i)q(\theta^*|\theta_i)} \right)
$$

	and $\theta_{i+1} = \theta_i$ with probability $1 - \alpha$. 
	
	\item [M6] Set $i = i+1$, go to \textbf{M2}.

\end{description}

The outcome of this algorithm is a Markov chain with stationary distribution $\pi(\theta|d(x_0, x^*) \leq \epsilon)$
\citep{Marjoram:2003p5}. That is, ABC MCMC is guaranteed to converge to the target approximate posterior distribution. Notice that if the proposal distribution is symmetric, $q(\theta_i|\theta^*) = q(\theta^*|\theta_i)$, then $\alpha$ depends only on the prior distribution. Further, if the prior is uniform, then $\alpha = 1$ in {\bf M5}. 
Potential disadvantages of the ABC MCMC algorithm are that the correlated nature of samples
coupled with the potentially low acceptance probability may result in very long chains and that the chain may get stuck in regions of low probability for long periods of time.

The above mentioned disadvantages of ABC rejection and ABC MCMC methods can, at least in parts, be avoided in ABC algorithms based on sequential Monte Carlo (SMC) methods, first developed by \cite{Sisson:2007p2}. In this paper we derive ABC SMC from a
sequential importance sampling (SIS) algorithm \citep{DelMoral:2006p4, DelMoral:2008p10336}; see Appendix A for the derivation and B for a comparison with the algorithm of \cite{Sisson:2007p2}.

In ABC SMC, a number of sampled parameter values (called particles), $\{ \theta^{(1)}, \ldots, \theta^{(N)} \}$,  sampled from the prior distribution, $\pi(\theta)$, is propagated through a sequence of intermediate distributions, $\pi(\theta|d(x_0, x^*) \leq \epsilon_i)$, $i = 1,\ldots,T-1$,
until it represents a sample from the target distribution, $\pi(\theta|d(x_0, x^*)\leq \epsilon_T)$. The tolerances, $\epsilon_i$, are chosen such that $\epsilon_1 > \ldots > \epsilon_T \geq 0$, thus the distributions gradually evolve towards the target posterior. 
For sufficiently large numbers of particles the population approach can also in principle avoid the problem of getting
stuck in areas of low probability encountered in ABC MCMC. The \textit{ABC SMC algorithm} proceeds as follows\footnote{For a more general version of the algorithm, suitable especially for application to stochastic models, see Appendix A.}:

 \begin{description}

 	\item [S1] Initialize $\epsilon_1, \ldots, \epsilon_T$.\\
Set the population indicator $t = 0$. 

	\item [S2.0] Set the particle indicator $i = 1$.

	\item [S2.1] If $t = 0$, sample $\theta^{**}$ independently from $\pi(\theta)$.\\
	Else, sample $\theta^*$ from the previous population $\{\theta_{t-1}^{(i)}\}$ with weights $w_{t-1}$ and perturb the particle to obtain $ \theta^{**} \sim K_t(\theta|\theta^{*})$, where $K_t$ is a perturbation kernel.\\ 
	If $\pi(\theta^{**}) = 0$, return to \textbf{S2.1}.\\
	Simulate a candidate dataset $x^{*} \sim f(x|\theta^{**})$.\\
	If $d(x^{*}, x_0) \geq \epsilon_t$, return to \textbf{S2.1}.

	\item [S2.2] Set $\theta_t^{(i)} = \theta^{**}$ and calculate the weight for particle $\theta^{(i)}_t$,
	$$
	w_t^{(i)} =
	\left\{ \begin{array}{ll}
	1,  &  \textrm{if } t=0,  \\
	\frac{\pi(\theta^{(i)}_t)}{ \sum_{j=1}^N w_{t-1}^{(j)} K_t(\theta_{t-1}^{(j)},\theta^{(i)}_t)}, & \textrm{if } t >0.
	\end{array} \right.
	$$
	If $i < N$ set $i = i+1$, go to \textbf{S2.1}.

	\item [S3] Normalize the weights.\\
	If $t<T$, set $t = t+1$, go to \textbf{S2.0}.   

\end{description}
Particles sampled from the previous distribution are denoted by a single asterisk, and after perturbation these particles are denoted by a double asterisk. Here we choose the perturbation kernel $K_t$ to be a random walk (uniform or Gaussian).
%
Note that for the special case when $T = 1$, the ABC SMC algorithm corresponds to the ABC rejection algorithm.

\subsection{Model selection}
Here we introduce an ABC SMC model selection framework which employs standard concepts from Bayesian model selection, including Bayes factors (a comprehensive review of Bayesian model selection can be found in \cite{Kass:1995p2898}). Let $m_1$ and $m_2$ be
two models; we would like to choose which model explains the data, $x$, better. The Bayes factor is defined as
\begin{equation} \label{BF}
B_{12} = \frac{P(m_1|x)/P(m_2|x)}{P(m_1)/P(m_2)},
\end{equation}
where $P(m_i)$ is the prior and $P(m_i|x)$ the posterior distribution of model $m_i$, $i = 1, 2$. If the priors are uniform, then (\ref{BF}) simplifies to 
\begin{equation} \label{BF_uniform}
B_{12} = \frac{P(m_1|x)}{P(m_2|x)}.
\end{equation}
The Bayes factor is a summary of the evidence provided by the data in favour of one statistical model over another (see Table \ref{tab:table1} for its interpretation).
\begin{table}
\caption{\small{Interpretation of Bayes factor, adapted from \cite{Kass:1995p2898}.}}
\label{tab:table1}
\begin{center}
\begin{tabular}{c|c}
	The value of & Evidence against $m_2$  \\
	Bayes factor $B_{12}$ & (and in favour of $m_1$) \\
	\hline
	$1$ to $3$ & Very weak \\
	$3$ to $20$ & Positive \\
	$20$ to $150$ & Strong \\
	$> 150$ & Very strong \\
\end{tabular}
\end{center}
\end{table}

There are several advantages of Bayesian model selection compared to traditional hypothesis testing. Firstly, the models being
compared do not need to be nested. Secondly, Bayes factors do not only weigh the evidence against a  hypothesis (in our
case $m_2$), but can equally well provide evidence in favour of it. This is not the case for traditional hypothesis testing where a small $p$-value only indicates that the
null model has insufficient explanatory power. However, one cannot conclude from a large $p$-value that the two models are equivalent, or that the
null model is superior, but only that there is not enough evidence to distinguish between them. In other words, ``failing to reject'' the null hypothesis cannot be translated into ``accepting'' the null hypothesis \citep{Cox:1974p15668,Kass:1995p2898}. Thirdly, unlike the posterior probability of the model, the $p$-value does not provide any direct interpretation of the weight of
evidence (the $p$-value is not the probability that the null hypothesis is true). 
\par
Here we approach the model selection problem by including a ``model parameter'' $m \in \{ 1,\ldots,M \}$, where $M$ is the number of models, as an additional discrete parameter and denote the model-specific parameters as $\theta(m) = (\theta(m)^{(1)}, \ldots, \theta(m)^{(k_m)})$, $m=1,\ldots,M$, where $k_m$
denotes the number of parameters in model $m$. 

In each population we start by sampling a model indicator $m$ from the prior distribution $\pi(m)$. For model $m$ we then propose new particles by perturbing the particles from the previous population specific to $m$; this step is the same as in the parameter estimation algorithm. The weights for particles $\theta(m)$ are also calculated like in the parameter estimation algorithm for $m$. 

The \textit{ABC SMC algorithm for model selection} proceeds as follows\footnote{In stochastic framework we again suggest using the general form of the algorithm with $B_t > 1$, see Appendix A.}:
\begin{description}

 	\item [MS1] Initialize $\epsilon_1, \ldots, \epsilon_T$.\\
Set the population indicator $t = 0$. 

	\item [MS2.0] Set the particle indicator $i = 1$.

	\item [MS2.1] Sample $m^{*}$ from $\pi(m)$. \\  
	If $t = 0$, sample $\theta^{**}$ from $\pi(\theta(m^{*}))$.\\
	If $t > 0$, sample $\theta^*$ from the previous population $\{ \theta(m^{*})_{t-1}\}$ with weights ${w(m^{*})_{t-1}}$.\\
	Perturb the particle $\theta^*$ to obtain $\theta^{**}\sim K_t(\theta|\theta^{*})$.\\	
	If $\pi(\theta^{**}) = 0$, return to \textbf{MS2.1}.\\
	Simulate a candidate dataset $x^{*} \sim f(x|\theta^{**},m^*)$.\\
	If $d(x^{*}, x_0) \geq \epsilon_t$, return to \textbf{MS2.1}.

	\item [MS2.2] Set $m_t^{(i)} = m^{*}$ and add $ \theta^{**}$ to the population of particles $\{ \theta(m^{*})_t\}$, and calculate its weight as	
	$$
	w_t^{(i)} =
	\left\{ \begin{array}{ll}
	1,  &  \textrm{if } t=0,  \\
	\frac{\pi(\theta^{**})}{ \sum_{j=1}^{N} w_{t-1}^{(j)} K_t(\theta_{t-1}^{(j)},\theta^{**}) }, & \textrm{if } t >0.
	\end{array} \right.
	$$	
	If $i < N$ set $i = i+1$, go to \textbf{MS2.1}.  
	
	\item [MS3] For every $m$, normalize the weights.\\
	If $t<T$, set $t = t+1$, go to \textbf{MS2.0}.   

\end{description}
The outputs of the ABC SMC algorithm are approximations of the marginal posterior distribution of the ``model parameter'' $P(m|x)$ and the marginal posterior distributions of parameters $P(\theta_i | x,m)$, $m=1,\ldots,M$, $i = 1,\ldots,k_m$.  Note that it can happen that a model dies out (\textit{i.e.} there are no particles left that belong to a particular model) if it offers only a poor description of the data. In this case the sampling of particles continues from the remaining models only.

Bayes factors can be obtained directly from $P(m|x)$ using equation (\ref{BF_uniform}). However, in many cases there will not be a single best and most
powerful/explanatory model \citep{Stumpf:2006p15704}. More commonly, different models explain different parts of the data to a certain extent.
One can average over these models to get a better inference than from a single model only. The approximation of the marginal posterior distribution of the model,
$P(m|x)$, which is the output of the above algorithm, can be used for Bayesian model averaging \citep{Hoeting:1999p3168}. 

The parameter estimation for each of the models is done simultaneously with the model selection. The
model with the highest posterior probability will typically  have the greatest number of particles, thereby ensuring a good estimate of the
posterior distribution for the parameters. However, some models are poorly represented in the marginal posterior distribution of $m$ (i.e. only a small number of particles belong to these models), and so the small number of particles does not provide very good estimated of the posterior distributions on parameters. Therefore, one might wish to estimate parameters for these models independently.

We note that the ABC SMC model selection algorithm implicitly penalizes the models with large number of parameters; the higher the parameter dimension is, the smaller is the probability that the perturbed particle is accepted.

\subsection{Implementation of the algorithm}
The algorithm is implemented in C++. For the ODE solver code the $4^{th}$ order classical Runge Kutta algorithm from the GSL Scientific Library \citep{Galassi:2003p15775} is used; for the simulation
of stochastic models we use Gillespie's algorithm \citep{Gillespie:1977p13997}; and for the simulation of delay differential equations we implemented the algorithm based on the
adaptive step size ODE solvers from Numerical recipes in C \citep{HPress:1992p15843} extended by code handling the delay part according to \cite{Paul:1992p16032} and
\cite{Enright:1995p16061}.

\section{Results}

We demonstrate the performance of the ABC algorithms using simulated data from deterministic and stochastic systems. The data points were obtained by solving the systems
for some fixed parameters at chosen time points. The sizes of the input datasets were chosen to reflect what can typically be expected in
real-world datasets in ecology, molecular systems biology and epidemiology. The first two examples highlight the computational performance of ABC SMC, the problem of inferability of dynamical models and its relationship to parameter sensitivity. The third example illustrates the use of ABC SMC for model selection, which is then further demonstrated in an application to a real dataset.

\subsection{Parameter inference for deterministic and stochastic Lotka-Volterra model}

The first model is the Lotka-Volterra (LV) model \citep{Lotka:1925p16151, Volterra:1926p16212} describing the interaction between prey species, $x$, and predator species, $y$, with parameter vector $\theta = (a,b)$:
\begin{subequations}
	\begin{eqnarray}
		\frac{dx}{dt} &=& ax - xy, \\
		\frac{dy}{dt} &=& bxy - y.
	\end{eqnarray}
\end{subequations}

\subsubsection{Computational efficiency of ABC SMC applied to deterministic LV dynamics}
\begin{figure}[thp]	
	\centering
	\includegraphics[width = 90mm]{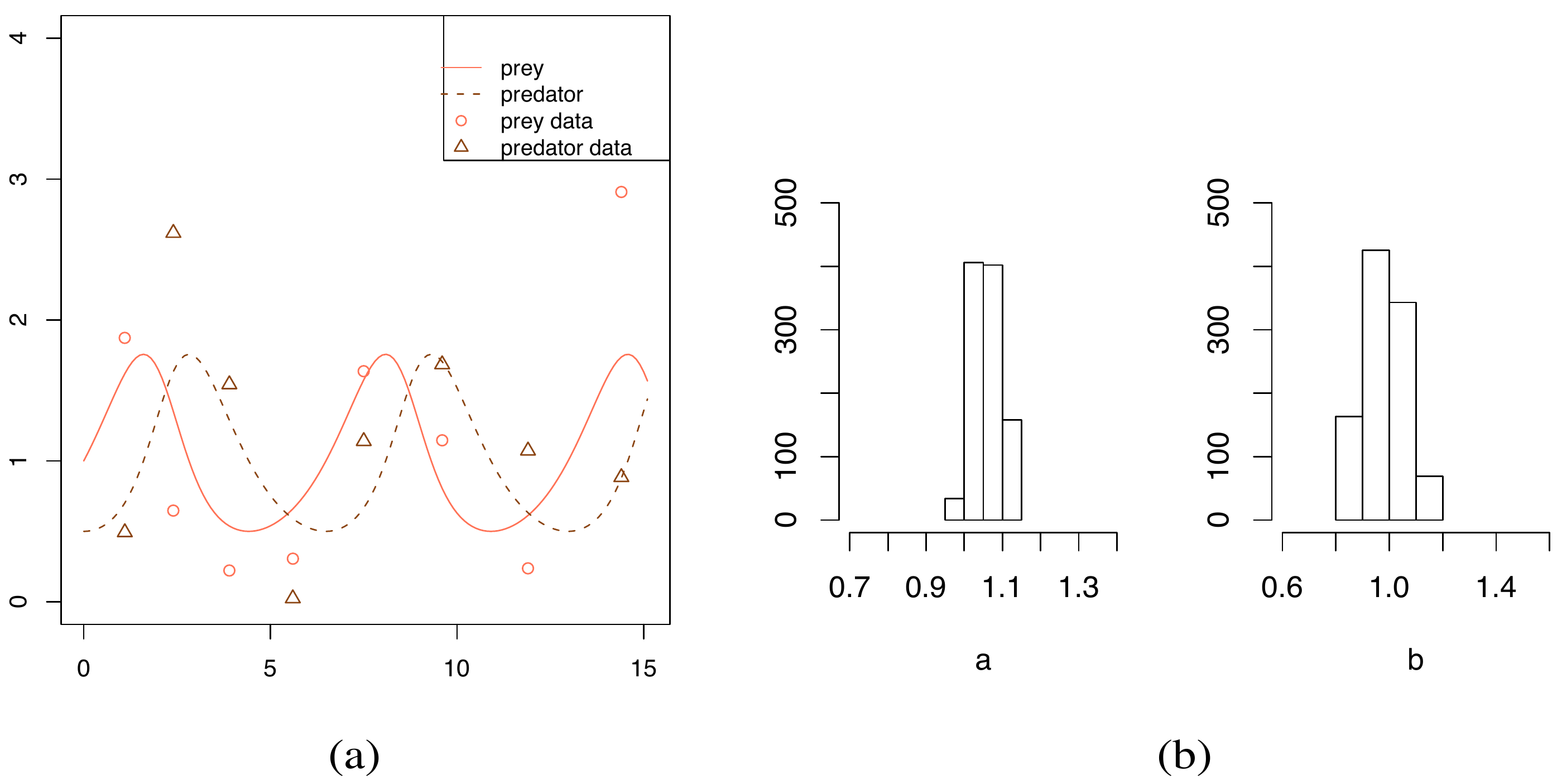} 
	\caption{\small{(a) Trajectories of prey and predator populations of the deterministic Lotka-Volterra system and the data points. (b) Parameters inferred by the ABC rejection sampler.}\label{joint_LV_graph}}
\end{figure}

The data, $\{(x_d,y_d)\}$, are noisy observations of the simulated system with parameter values set at $(a,b) = (1,1)$. We sample $8$ data points (for each of the species) from the solution of the system for parameters $(a,b)=(1,1)$ and add Gaussian noise ${\cal
N}(0,(0.5)^2)$ (see Figure \ref{joint_LV_graph}(a)). Let the distance function $d((x_d,y_d),(x,y))$ between the data $\{(x_d[i],y_d[i])\}$, $i = 1, \ldots, 8$ and a simulated solution for proposed parameters, $\{(x[i],y[i])\}$, 
be the sum of squared errors:

\begin{equation} \label{squared_errors}
d((x,y),(x_d, y_d)) = \sum_i\left( (x[i]-x_d[i])^2 + (y[i]-y_d[i])^2 \right).
\end{equation}
In Appendix C we show that this distance function is, in fact, related to the conventional likelihood treatment of ODEs.
\par 
The distance between our noisy data and the deterministic solution for $(a,b) = (1,1)$ from which the data was generated is $4.23$, so the lowest distance to be reached is expected to be close to this number and we choose the tolerance $\epsilon$ accordingly.

We first apply the ABC rejection sampler approach with $\epsilon = 4.3$. The prior distributions for $a$ and $b$ are taken to be uniform,
$a,b \sim U(-10, 10)$. In order to get $1,000$ accepted
particles, approximately $14.1$ million data generation steps are needed, which means that the acceptance rate ($7 \cdot 10^{-5}$) is
extremely low. The inferred posterior distributions are shown in Figure \ref{joint_LV_graph}(b).

Applying the ABC MCMC scheme outlined above yields results comparable to those of ABC rejection, and after a careful calibration of the approach (using an adaptive Gaussian proposal distribution) we manage to markedly reduce the computational cost (including burn-in, we had to generate between $40,000$ and $60,000$ simulations in order to obtain $1,000$ independent particles). 

\begin{figure}[h]	
	\begin{center}
	\includegraphics[width = 77mm]{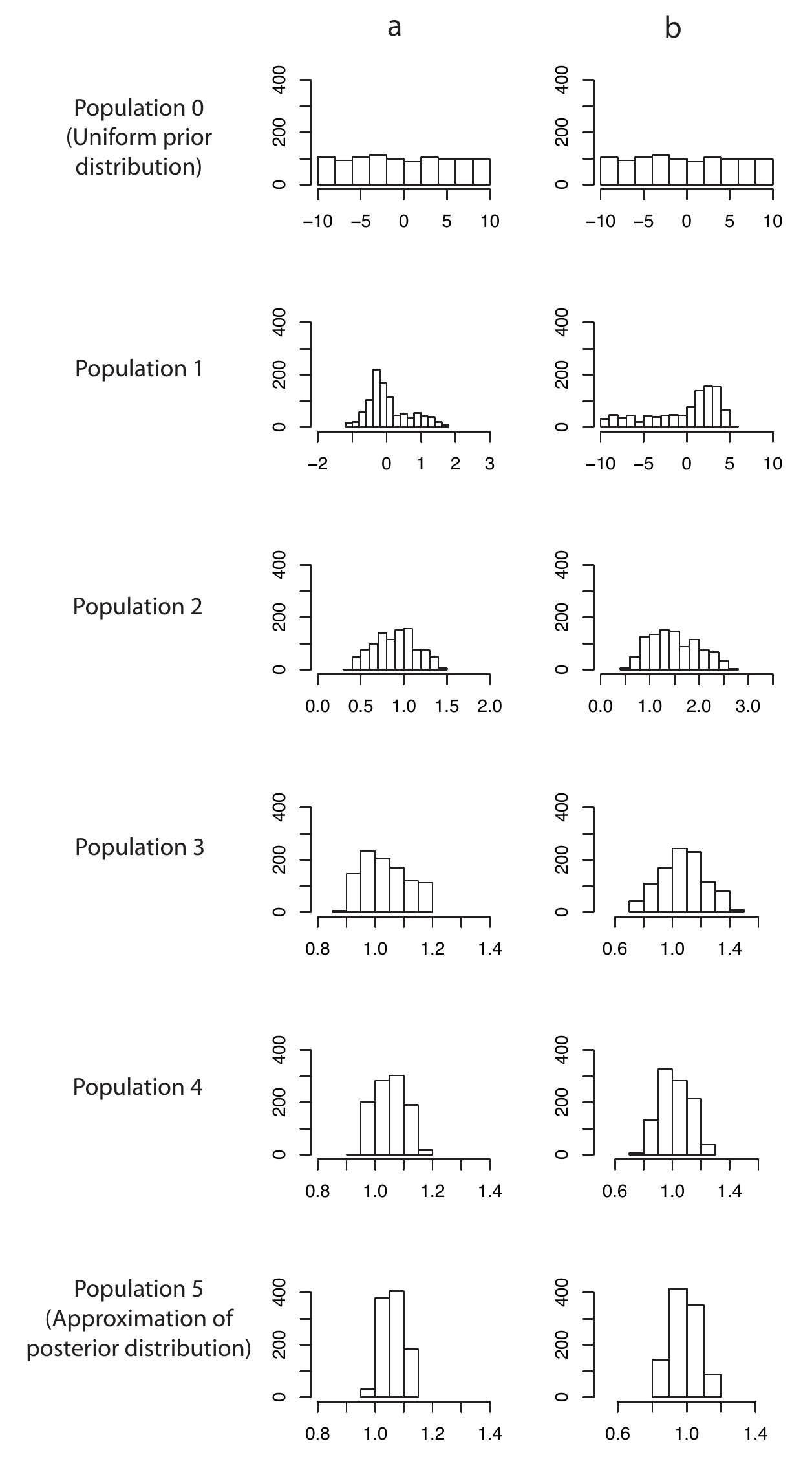}
	\end{center}
	\caption{\label{LV_ODE}\small{Histograms of populations $0$ -- $5$ of parameters $a$ and $b$ of the Lotka-Volterra system.}}
\end{figure}

We next apply the ABC SMC approach. The prior distributions for $a$ and $b$ are taken to be uniform,
$ a,b \sim U(-10, 10)$, and the perturbation kernels for both parameters are uniform, $K_t = \sigma U(-1,1)$, with $\sigma = 0.1$. The number of particles in each population is $N = 1000$. To ensure the gradual transition between populations we take $T=5$ populations with $\epsilon = (30.0, 16.0, 6.0, 5.0, 4.3)$. The results are summarized in Table \ref{tab:table2} and Figure
\ref{LV_ODE}. From the last population (population 5) it can be seen that both
parameters are  well inferred (a: median = $1.05$, 95\%-quantile range = $[1.00, 1.12]$, b:
median = $1.00$, 95\%-quantile range = $[0.87, 1.11]$). The outcome is virtually the same as previously obtained
by the ABC rejection sampler (Figure \ref{joint_LV_graph}(b)), however, there is a substantial reduction in the number
of steps needed to reach this result. For this model the ABC SMC algorithm needs $50$-times fewer data generation steps than the ABC rejection sampler, and about the same number of data generation steps than the adaptive ABC MCMC sampler.  

\begin{table}
\caption{\small{Cumulative number of data generation steps needed to accept 1000 particles in each population for deterministic LV dynamics.}}
\label{tab:table2}
\begin{center}
\begin{tabular}{c|ccccc}
	Population & 1 & 2 &  3 &  4 &  5 \\ 
	\hline
	Data gen. steps & 26,228 & 36,667 & 46,989 & 49,271 & 52,194
\end{tabular}
\end{center}
\end{table}

The analyses were repeated with different distance functions, such as
\begin{equation} \label{L1_norm}
d((x,y),(x_d, y_d)) = \sum_{i}\left( |x[i]-x_d[i]| + |y[i]-y_d[i]| \right)
\end{equation}
and 
\begin{equation} \label{2_norm}
d((x,y),(x_d, y_d)) = 2 - \frac{x \cdot x_d}{||x||||x_d||} - \frac{y \cdot y_d}{||y||||y_d||}, 
\end{equation}
where ``$\cdot$'' denotes the inner product. As expected, the resulting approximations of posterior distributions are very similar (histograms not shown). Replacing the uniform perturbation kernel by a Gaussian kernel also yields the same results, but requires more simulation steps (results not shown). 


\subsubsection{ABC SMC inference for stochastic LV dynamics}
Having obtained good estimates for the deterministic case, we next try to infer parameters of a stochastic LV model. The predator-prey process can be described by the following rate equations:
\begin{subequations}
	\label{LV_stochastic}
	\begin{eqnarray}
		a + X &\rightarrow& 2X  \hspace{1cm} \textrm{with rate } c_1,\\
		X+Y &\rightarrow& 2Y  \hspace{1cm} \textrm{with rate } c_2,\\
		Y &\rightarrow& \emptyset  \hspace{1.35cm} \textrm{with rate } c_3,
	\end{eqnarray}
\end{subequations}
where $X$ denotes the prey population, $Y$ the predator population and $a$ the fixed amount of resource available to the prey (we fix it to $a=1$).
These reactions define the stochastic master equation \citep{vanKampen:2007p16760}
\begin{eqnarray} \label{LV_master equation}
\frac{\partial P(x,y,t)}{\partial t} &=& c_1 a(x-1)P(x-1,y,t) \nonumber \\
&+& c_2 (x+1)(y-1)P(x+1,y-1,t) \nonumber \\
&+& c_3 (y+1) P(x,y+1,t) \nonumber \\
&-& (c_1 ax + c_2 xy + c_3 y)P(x,y,t).
\end{eqnarray}
The ABC approach can easily be applied to inference problems involving master equations, because there exists an exact simulation algorithm (Gillespie algorithm \citep{Gillespie:1977p13997,Wilkinson:2006p2403}). 

Our simulated data consists of $19$ data points for each species with rates $(c_1,
c_2, c_3) = (10, 0.01, 10)$, and initial conditions $(X_0, Y_0) = (1000,1000)$. The distance function is the sum of squared
errors (\ref{squared_errors}) and the SMC algorithm is run for $T=5$ populations with $\epsilon = (4000, 2900, 2000, 1900, 1800)$. Especially in laboratory settings, results from several replicate experiments  are averaged over; here we therefore also use data averaged over three independent replicates. The simulated data at every run are then, like the experimental data, averaged over three runs. 
For inference purposes the average over several runs tends to hold more information about the system's mean dynamics then a single stochastic run. If the experimental data consisted of one run only, \textit{i.e.} if there were no repetitions, then the inference could in principle proceed in the same way, by comparing the experimental data to a single simulated run only. This would result in the lower acceptance rate and consequently more data generation steps to complete the inference.

\begin{figure}[h]	

	\begin{center}
	\includegraphics[width = 80mm]{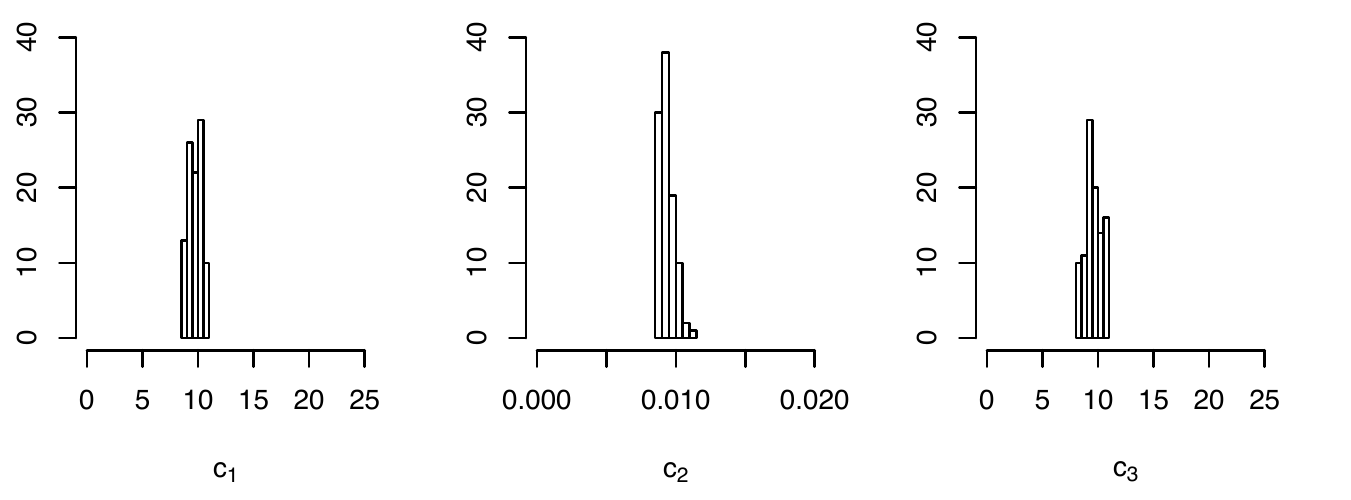}
	\end{center}
	\caption{\label{fig:LV_stochastic}\small{Histograms of the approximated posterior distributions of parameters $c_1$, $c_2$ and $c_3$ of the stochastic LV dynamics.}}
\end{figure}

We generate $N = 100$ particles per population and assign the prior distributions on the parameters to be $\pi(c_1) = U(0,28)$, $\pi(c_2) = U(0.0, 0.04)$ and $\pi(c_3) =
U(0,28)$, reflecting the respective orders of magnitude of the simulated parameters (if a larger domain for $\pi(c_2)$ is taken, there are no accepted instances for
$c_2>0.04$). Perturbation kernels are uniform with $\sigma_{c_1} = \sigma_{c_3} = 1.0$ and $\sigma_{c_2} = 0.0025$, and $B_t = 10$. The results are summarized in 
Figure \ref{fig:LV_stochastic}. 
%

\subsection{Parameter inference for the deterministic and stochastic repressilator model}

The repressilator \citep{Elowitz:2000p13} is a popular toy model for gene-regulatory systems and consists of three genes connected in a feedback loop, where each gene transcribes the repressor protein for the next gene in the loop and
is repressed by the previous gene. The model consists of six ordinary differential equations and four parameters: 
\begin{subequations} \label{eqn:rep_deterministic}
	\begin{eqnarray}
		\frac{dm_1}{dt} &=& -m_1 + \frac{\alpha}{1+p_3^n} + \alpha_0 \\
		\frac{dp_1}{dt} &=& -\beta (p_1 - m_1) \\
		\frac{dm_2}{dt} &=& -m_2 + \frac{\alpha}{1+p_1^n} + \alpha_0 \\
		\frac{dp_2}{dt} &=& -\beta (p_2 - m_2) \\
		\frac{dm_3}{dt} &=& -m_3 + \frac{\alpha}{1+p_2^n} + \alpha_0 \\
		\frac{dp_3}{dt} &=& -\beta (p_3 - m_3)
	\end{eqnarray}
\end{subequations}

\subsubsection{Inferability and Sensitivity in deterministic repressilator dynamics}
Let $\theta = (\alpha_0, n, \beta, \alpha)$ be the parameter vector. For the simulated data the initial conditions are $(m_1, p_1, m_2, p_2, m_3, p_3) = (0,2,0,1,0,3)$ and the values of parameters are $\theta = (1, 2, 5, 1000)$; for these parameters the repressilator displays limit-cycle behaviour. We assume that only the mRNA $(m_1, m_2, m_3)$ data measurements are available and protein concentrations are considered as the missing data. Gaussian noise ${\cal N}(0,5^2)$ is added to the data points. The distance function is defined to be the squared errors. The prior parameter distributions are chosen as follows: $\pi(\alpha_0) = U(-2,10)$, $\pi(n) = U(0,10)$, $\pi(\beta) =
U(-5,20)$ and $\pi(\alpha) = U(500,2500)$. We assume the initial conditions are known. 

\begin{figure}[thp]	
	\centering
         \includegraphics[width = 135mm]{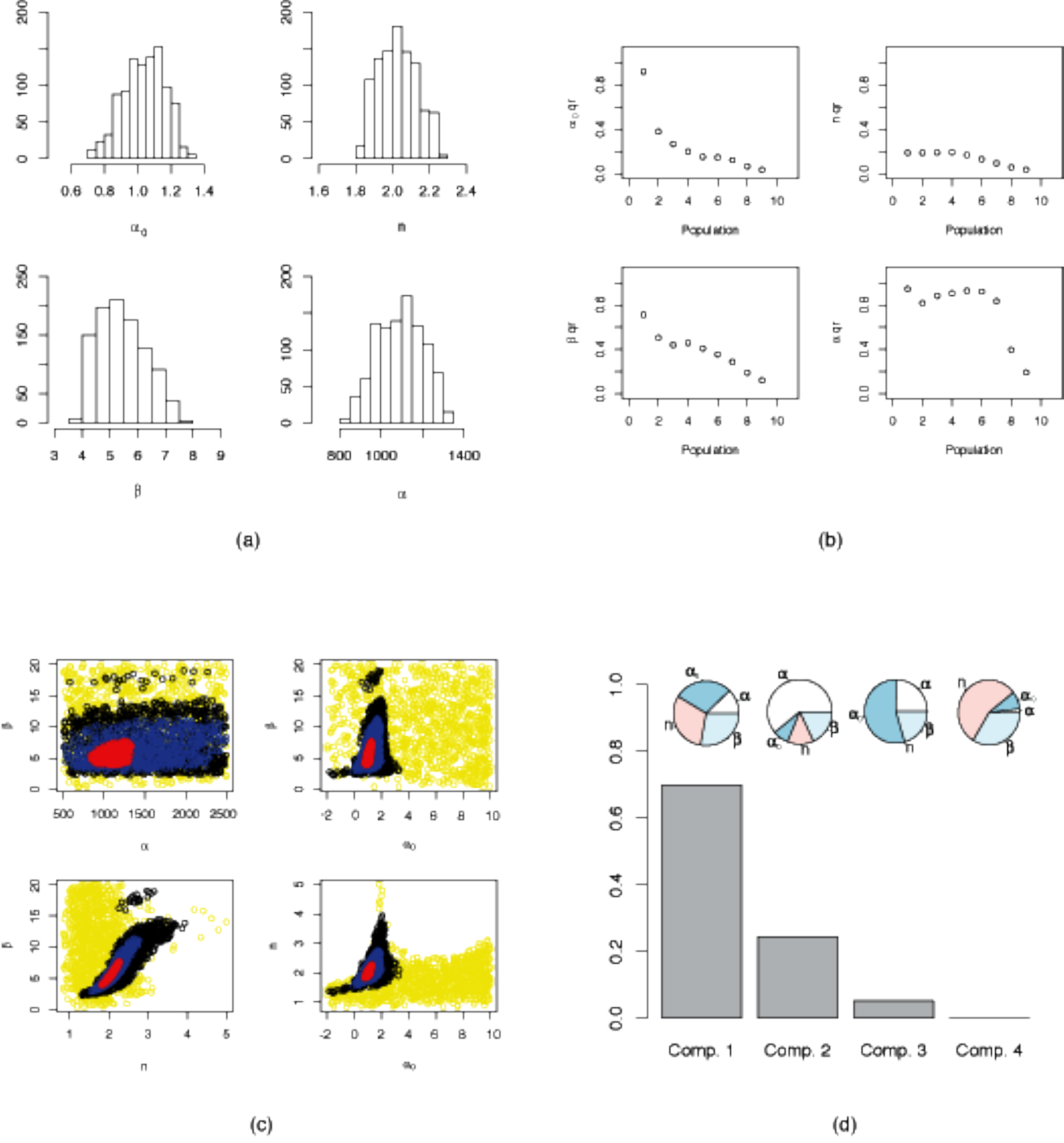}
	\caption{\small{
	(a) Histograms of the approximate marginal posterior distributions of parameters $\alpha_0$, $n$, $\beta$ and $\alpha$ of the deterministic repressilator model.
	(b) The normalized $95\%$ inter-quantile ranges of each population. The narrower the interval for a given tolerance $\epsilon_t$, the more sensitive the model is to the corresponding parameter. The interquartile range reached in population $9$ is determined by the added experimental noise. As $\epsilon_9$ was chosen accordingly, one cannot proceed by lowering the tolerance further. The sharp change in the interquartile ranges, which occurs, for example, for parameter $\alpha_0$ between populations $0$ and $1$, can be explained by the steep gradient of the likelihood surface along $\alpha_0$.
	(c) The output (i.e. the accepted particles) of ABC SMC algorithm as two dimensional scatterplots.  The particles from population $1$ are coloured in yellow, particles from population $4$ in black, particles from population $7$ in blue, and particles from the last population in red. Islands of particles are observed in population $4$ and they can be explained by the multi-modality of the $4^{th}$ intermediate distribution.
	(d) PCA of the set of accepted particles (population $9$). Due to dependence of PCA to the scaling of original variables the PCA was done on the correlation matrix. The first PC explains $70.0\%$ of the total variance, the second $24.6\%$, the third $5.3\%$ and the fourth $0.1\%$ of the variance. Pie charts show the fraction of the length of PCs explained by individual parameters. } \label{fig:rep_determ_joint}}
\end{figure}

The results are summarized in Figures \ref{fig:rep_determ_joint}(a) - \ref{fig:rep_determ_joint}(d), where we show the posterior distributions, the change in inter-quartile ranges of parameter estimates across the populations and scatterplots of some of the two-dimensional parameter combinations.
%
%
Each parameter is easily inferred when the other three parameters are  fixed (histograms not shown). When the algorithm is applied to all four parameters simultaneously, parameter $n$ is inferred the quickest and has the smallest posterior variance, while parameter $\alpha$ is barely inferable and it has large credible intervals (Figures \ref{fig:rep_determ_joint}(a) and \ref{fig:rep_determ_joint}(c)). 
\par
We find that ABC SMC recovers the intricate link between model sensitivity to parameter changes and inferability of parameters. The repressilator system is most sensitive to changes in parameter $n$ and least to changes in $\alpha$. Hence, data appears to contain little information about $\alpha$. Thus ABC SMC provides us with a global parameter sensitivity analysis \citep{Sanchez:1997p16286} on the fly as intermediate distributions are being constructed. Note that the intermediate distributions are nested in one another (as should be the case in SMC) (Figure \ref{fig:rep_determ_joint}(c)).  An attempt to visualize the four-dimensional posterior distributions can be accessed in the supplementary material where we provide an animation in which the posterior distribution in four-dimensional parameter space is projected onto two dimensional planes. 

The interquartile ranges and the scatterplots provide an initial impression of parameter sensitivity, however, the first problem with scatterplots is that it is increasingly difficult to visualize the behaviour of the model with increasing parameter dimension. Secondly, we have to determine the sensitivity when a combination of parameters is varied (and not just individual parameters), and this cannot be visualized via simple one-dimensional interquartile ranges or two-dimensional scatterplots. 

We can use principal component analysis (PCA) to quantify the  sensitivity of the system \citep{Saltelli:2008p14297}. The output of the ABC SMC algorithm, which we are going to use for our sensitivity analysis, is the last population of $N$ particles. Associated with the accepted particles is their variance-covariance matrix, ${\Sigma}$, of rank $p$, where $p$ denotes the dimension of the parameter vector. The principal components (PC) are the eigenvectors of $\Sigma$ which define a set of eigen-parameters, $\chi_i=a_{i1}\theta_1+\dots+a_{ip} \theta_p$. Here  ${a}_i=(a_{i1},\ldots,a_{ip})$  is the normalized eigenvector associated with the $i^{th}$ eigenvalue of $\Sigma$, $\lambda_i$, and $a_{ij}$ describes the projection of parameter $\theta_j$ onto the $i^{th}$ eigen-parameter. PCA provides an orthogonal transformation of the original parameter space and the PCs can be taken  to define a $p$-dimensional ellipsoid which approximates the population of data points (\ie the accepted particles), and the eigenvalues specify the $p$ corresponding radii. The variance of the $i^{th}$ PC is given by $\lambda_i$ and the total variance of all PCs equals $\sum_{i=1}^{p}\lambda_i = \textrm{trace}(\Sigma)$. Therefore, the eigenvalue $\lambda_i$ associated with the $i^{th}$ PC explains a proportion 
\begin{equation}
\frac{\lambda_i}{\text{trace}({\Sigma})}
\end{equation}
of the variation in the population of points. The smaller the $\lambda_i$, the more sensitive the system is to variation of the eigen-parameter $\chi_i$. 
PCA only yields an approximate account of sensitivity similar to what would be obtained by computing the Hessian around the mode of the posterior distribution.  

Figure \ref{fig:rep_determ_joint}(d) summarizes the output of PCA. Is shows how much of a variance is explained by each PC, and which parameters contribute the most to these PCs. On the contrary to the interest in the first PC in most PCA applications, our main interest lies in the smallest PC. The last PC extends across the narrowest region of posterior parameter distribution, and therefore provides information on parameters to which the model is the most sensitive. In other words, the smaller PCs correspond to stiff parameter combinations, while the larger PCs may correspond to sloppy parameter combinations \citep{Gutenkunst:2007p728}.

%
%
%
The analysis reveals that the last PC mainly extends in the direction of a linear combination of parameters $n$ and $\beta$, from which we can conclude that the model is most sensitive to changes in these two parameters. Looking at the third component, the model is somewhat less sensitive to variation in $\alpha_0$. The model is therefore the least sensitive to changes in parameter $\alpha$, which is also supported by the composition of the $2^\text{nd}$ PC. This outcome agrees with the information obtained from the interquartile ranges and the scatterplots.

\subsubsection{Inference of stochastic repressilator dynamics}
We next apply ABC SMC to the stochastic repressilator model. We transformed the deterministic model (\ref{eqn:rep_deterministic}) into a set of the following reactions:
\begin{subequations}
	\label{rep_stochastic}
	\begin{eqnarray}
		\emptyset &\rightarrow& m_i  \hspace{1.15cm} \textrm{with hazard } \frac{\alpha}{1+p_j^n} + \alpha_0,\\
		m_i &\rightarrow& \emptyset  \hspace{1.4cm} \textrm{with hazard } m_i,\\
		m_i &\rightarrow& m_i+ p_i  \hspace{0.4cm} \textrm{with hazard } \beta m_i,\\
		p_i &\rightarrow& \emptyset  \hspace{1.4cm} \textrm{with hazard } \beta p_i,
	\end{eqnarray}
\end{subequations}
where $i = 1,2,3$ and, correspondingly, $j = 3,1,2$. The stochastic process defined by these reactions can be simulated with Gillespie algorithm. True parameters and initial conditions correspond to those of the deterministic case discussed above. The data includes both mRNA and protein levels at $19$ time points. Tolerances are chosen as $\epsilon = \{900,650,500,450,400\}$, the number of particles $N=200$ and $B_t = 5$. 

\begin{figure}[h]	
	\centering
	\includegraphics[width=140mm]{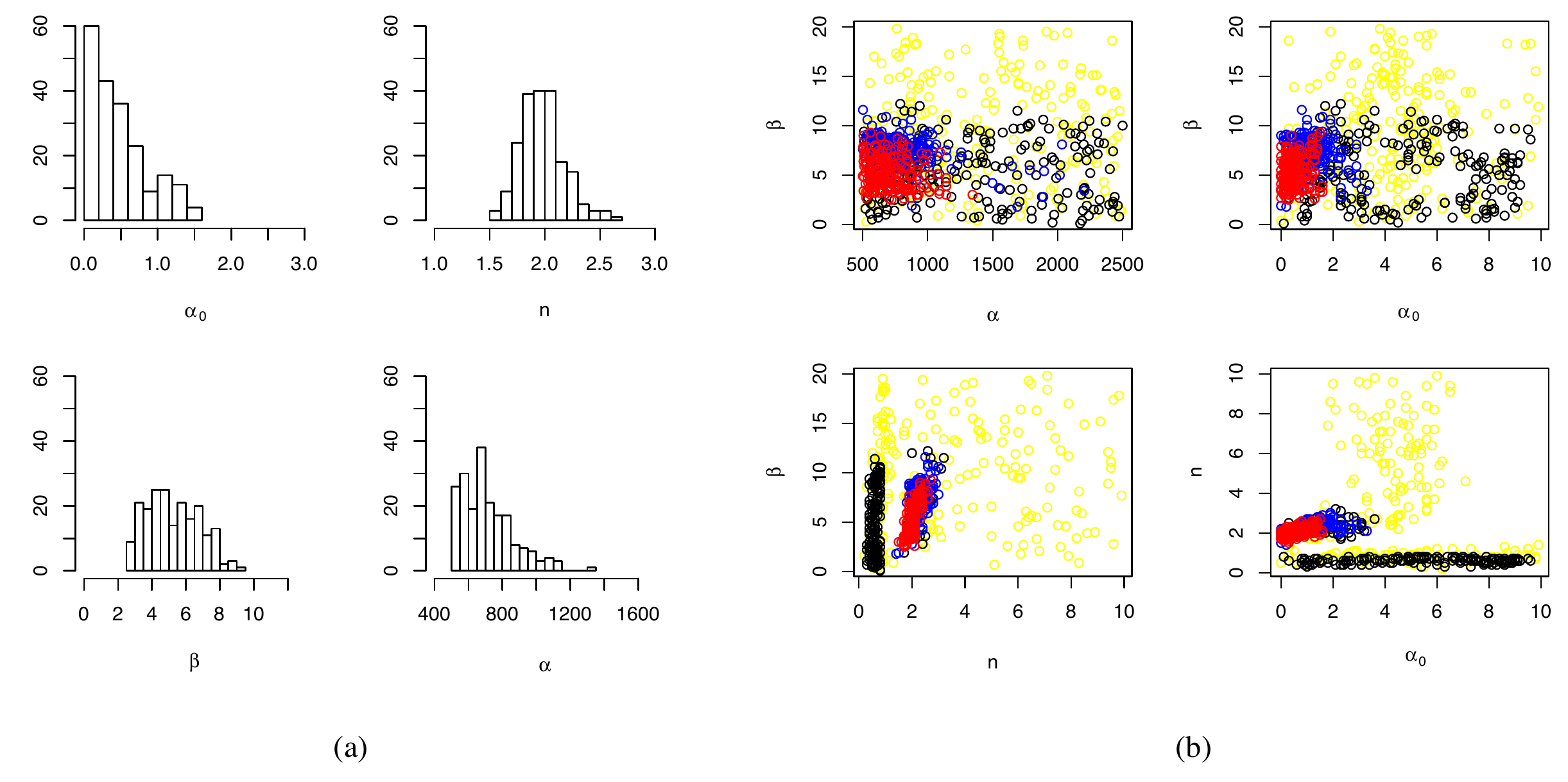}
	\caption{\small{(a) Histograms of the approximated posterior distributions of parameters $\alpha_0$, $n$, $\beta$ and $\alpha$ of the stochastic repressilator model. (b) The output of the ABC SMC algorithm as two dimensional scatterplots. The particles from population $1$ are coloured in yellow, particles from population $2$ in black, particles from population $3$ in blue, and particles from the last population in red. We notice that the projection to parameter $n$ in population $2$ is a sample from a multimodal intermediate distribution, while this distribution becomes unimodal from population $3$ onwards. }\label{fig:rep_stochastic_joint}}
\end{figure}

The inference results for comparing the average over 20 simulations with the data generated from the average of 20 simulations are summarized in Figures \ref{fig:rep_stochastic_joint}(a) and \ref{fig:rep_stochastic_joint}(b). The figures show that parameters $n$ and $\beta$ get reasonably well inferred, while $\alpha_0$ and $\alpha$ are harder to infer. It is clearly noticeable that parameters are better inferred in the deterministic case (Figure \ref{fig:rep_determ_joint}(a)). 

\subsubsection{Contrasting inferability for deterministic and stochastic dynamics}
Analyzing and comparing the results of the deterministic and stochastic repressilator dynamics shows that parameter sensitivity is intimately linked to inferability. If the system is insensitive to a parameter, then this parameter will be hard (or even impossible) to infer, as varying such a parameter does not vary the output -- which here is the approximate posterior probability -- very much. In stochastic problems we may furthermore have the scenario where the fluctuations due to small variations in one parameter overwhelm signals from other parameters. This seems to be the case here, where $\alpha_0$ is harder to infer in the stochastic case compared to the deterministic repressilator dynamics.

\subsection{Model selection on different SIR models}
We illustrate model selection using a range of simple models that can describe the epidemiology of infectious diseases. 
SIR models describe the spread of such disease in a population of susceptible (S), infected (I) and recovered (R) individuals \citep{Anderson:1991p16708}, respectively. The simplest model assumes that every
individual can only be infected once and that there is no time delay between the individual getting infected and their ability to infect other susceptible individuals:
\begin{subequations} \label{basic}
	\begin{eqnarray} 
		\dot{S} &=& \alpha - \gamma S I - dS,\\
		\dot{I} &=& \gamma S I - vI - dI,\\
		\dot{R} &=& vI - dR,
	\end{eqnarray}
\end{subequations}
where $\dot{x}$ denotes the time derivative of $x$, $\frac{dx}{dt}$. Individuals, who are born at rate $\alpha$, are susceptible; the death rate (irrespective of disease class, $S$, $I$ or $R$) , is $d$; the infection rate is denoted by $\gamma$ and the recovery rate by
$v$.

The model can be made more realistic by adding a time delay $\tau$ between the time an individual gets infected and the time when they become infectious,
\begin{subequations} \label{delay}
	\begin{eqnarray}
		\dot{S} &=& \alpha - \gamma S I(t-\tau) - dS,\\
		\dot{I} &=& \gamma S I (t-\tau)- vI - dI,\\
		\dot{R}&=& vI - dR.
	\end{eqnarray}
\end{subequations}

Another way of incorporating the time delay into the model is by including a population of individuals in a latent (L) phase of infection; in this state they are
infected but cannot yet infect others. The equations then become
\begin{subequations} \label{latent}
	\begin{eqnarray}
		\dot{S} &=& \alpha - \gamma S I - dS,\\
		\dot{L} &=& \gamma S I - \delta L - dL,\\
		\dot{I} &=& \delta L - vI - dI,\\
		\dot{R} &=& vI - dR.
	\end{eqnarray}
\end{subequations}
Here $\delta$ denotes the transition rate from the latent to the infective stage. 

Another extension of the basic model (\ref{basic}) allows the recovered individuals to become susceptible again (with rate $e$):
\begin{subequations} \label{thirdmodel}
	\begin{eqnarray} 
		\dot{S} &=& \alpha - \gamma S I - dS + eR,\\
		\dot{I} &=& \gamma S I - vI - dI,\\
		\dot{R} &=& vI - (d+e)R.
	\end{eqnarray}
\end{subequations}
 
There are obviously many more ways of extending the basic model, but here we restrict ourselves to the four models described above. Given the same initial conditions, the outputs of all models are very similar, which makes it impossible to choose the right
model by visual inspection of the data alone. Therefore some more sophisticated, statistically based methods needs to be applied for selecting the
best available model. 
Therefore we apply the ABC SMC algorithm for model selection, developed in section 2.2. We define a model parameter $m=1,2,3,4$, representing the above models in the same
order, and model-specific parameter vectors $\theta(m)$: $\theta(1) = (\alpha, \gamma, d, v)$, $\theta(2) = (\alpha, \gamma, d, v, \tau)$,
$\theta(3) = (\alpha, \gamma, d, v, \delta)$ and $\theta(4) = (\alpha, \gamma, d, v, e)$. 
\par
\begin{figure}[h]	
	\begin{center}
	\includegraphics[width=80mm]{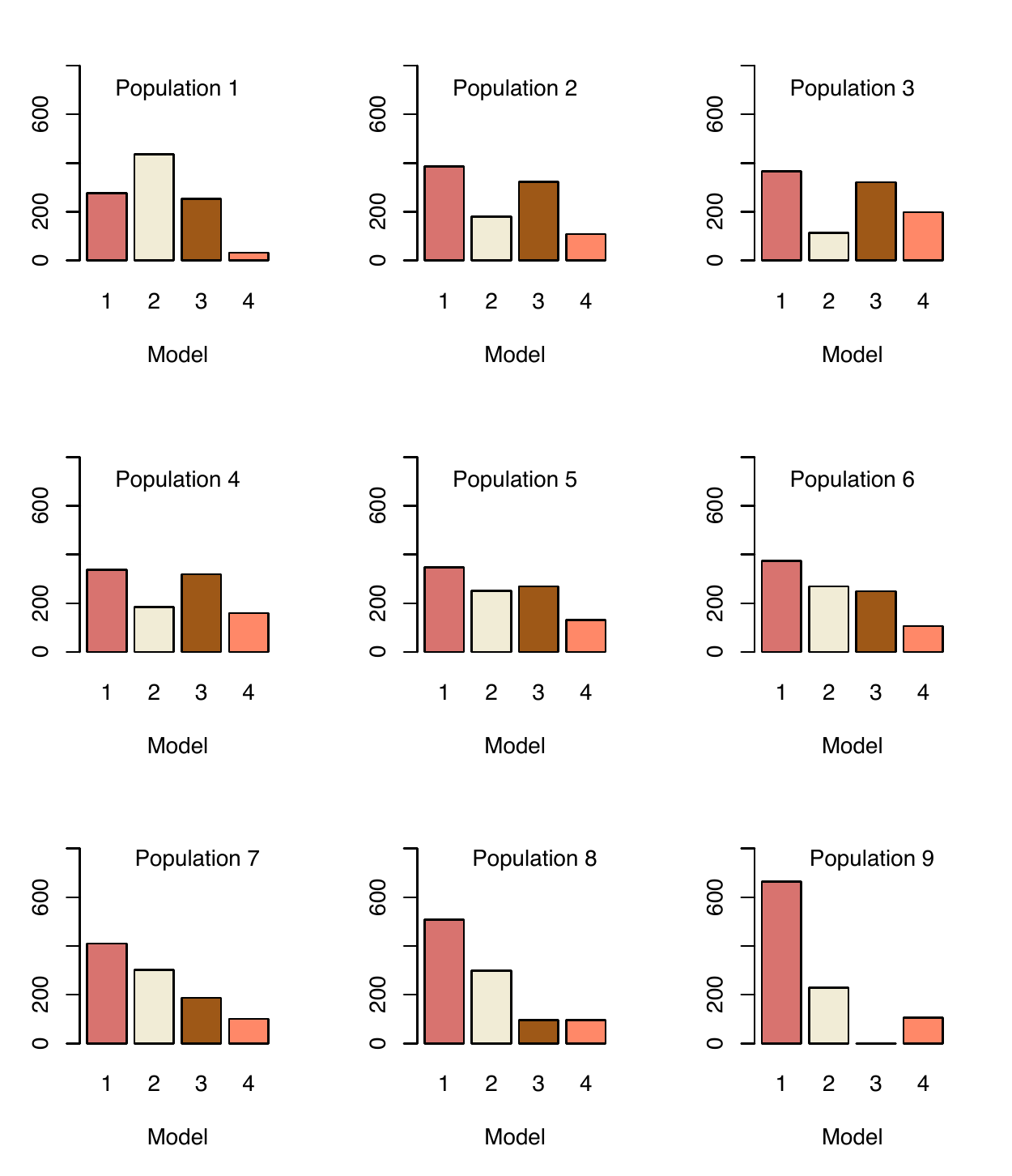}
	\end{center}
	\caption{\label{model_selection}\small{Histograms show populations $1$ -- $9$ of the model parameter $m$. Population $9$ represents the final posterior marginal estimate of $m$. Population $0$ (discrete uniform prior distribution) is not shown.}}
\end{figure}

 
Experimental data consists of $12$ data points from each of the three groups ($S$, $I$ and $R$). If the data are very noisy (Gaussian noise with standard deviation $\sigma = 1$ was added to simulated data points), then the algorithm cannot detect a single best model, which is not surprising given the high similarity of model outputs. However, if intermediate noise is added (Gaussian noise with standard deviation $\sigma = 0.2$), then the algorithm produces a posterior estimate with the most weight on the correct model. An example is shown in Figure \ref{model_selection}, where experimental data was obtained from model $1$ and perturbed by Gaussian noise, ${\cal N}(0,(0.2)^2)$. Parameter inference is done simultaneously with the model selection (posterior
parameter distributions not shown).

The Bayes factor can be calculated from the marginal posterior distribution of $m$ which we take from the final population.
From $1000$ particles model 1 (basic model) was selected $664$ times, model 2
$230$ times, model 4 $106$ times, and model 3 was not selected at all in the final population. Therefore, we can conclude from the Bayes factors
\begin{eqnarray}
B_{1,2} &=& \frac{664}{230} = 2.9,\\
B_{1,4} &=& \frac{664}{106} = 6.3,\\
B_{2,4} &=& \frac{230}{106} = 2.2,  
\end{eqnarray}
that there is weak evidence in favour of model 1 compared to model 2 and positive evidence in favour of model 1 compared to model 4. Increasing the amount of data will, however, change the Bayes factors in favour of the true model.

\subsection{Application: Common cold outbreaks on Tristan da Cunha}
\begin{table}
\caption{\small{Common cold data from Tristan de Cunha collected in October 1967.}}
\label{tab:tristan} 
\begin{tabular}{l|lllllllllllllllllllll}
	Day & 1 & 2  & 3 & 4 & 5 & 6 & 7 & 8  & 9  & 10 & 11 & 12  & 13 & 14 & 15 & 16 & 17 & 18  & 19  & 20 & 21\\
	\hline
	I(t) & 1 & 1  & 3 &  7 & 6 & 10 & 13 & 13  & 14  & 14 & 17 & 10  & 6 & 6 & 4 & 3 & 1 & 1  & 1  & 1 & 0\\	
	R(t) & 0 & 0  & 0 &  0 & 5 & 7 & 8 & 13  & 13  & 16 & 16 & 24  & 30 & 31 & 33 & 34 & 36 & 36  & 36  & 36 & 37\\
\end{tabular}
\end{table}
%

Tristan da Cunha is an isolated island in the Atlantic ocean with approximately $300$ inhabitants,
and it was observed that viral diseases, such as common cold, break out on the island after the arrival of ships from Cape Town. We use the $21$-day common cold data from
October 1967. The data, shown in Table \ref{tab:tristan}, was obtained from
\citep{Shibli:1971p16459,Hammond:1971p16458} and was previously used for parameter inference of a stochastic model by ABC MCMC  \citep{Cook:2008p16606}.
The data only provides the numbers of infected and recovered individuals, $I(t)$ and $R(t)$, whereas the size of initial susceptible population,
$S(0)$, is not known. Therefore, $S(0)$ is an extra unknown parameter to be estimated.

The four epidemic models from the previous section are used and because there are no births and deaths expected in the short period
of $21$ days, parameters $\alpha$ and $d$ are set to $0$. The tolerances are set to $\epsilon = \{ 100, 90, 80, 73, 70, 60, 50, 40, 30, 25, 20, 16, 15, 14, 13.8\}$
and $1000$ particles are used. The prior distributions of
parameters are chosen as follows: $\gamma \sim U(0,3)$, $v \sim U(0,3)$, $\tau \sim U(-0.5,5)$, $\delta \sim U(-0.5,5)$, $e \sim U(-0.5,5)$, $S(0)\sim U(37,100)$ and $m\sim U(1,4)$,
where $S(0)$ and $m$ are discrete. Perturbation kernels are uniform, $K_t = \sigma U(-1,1)$, with $\sigma_{\gamma} = \sigma_{v} = 0.3$, $\sigma_{\tau} = \sigma_{\delta} = \sigma_{e} = 1.0$ and $\sigma_{S(0)} = 3$.

\begin{figure}[hp]
	\begin{center}
	\includegraphics[width=80mm]{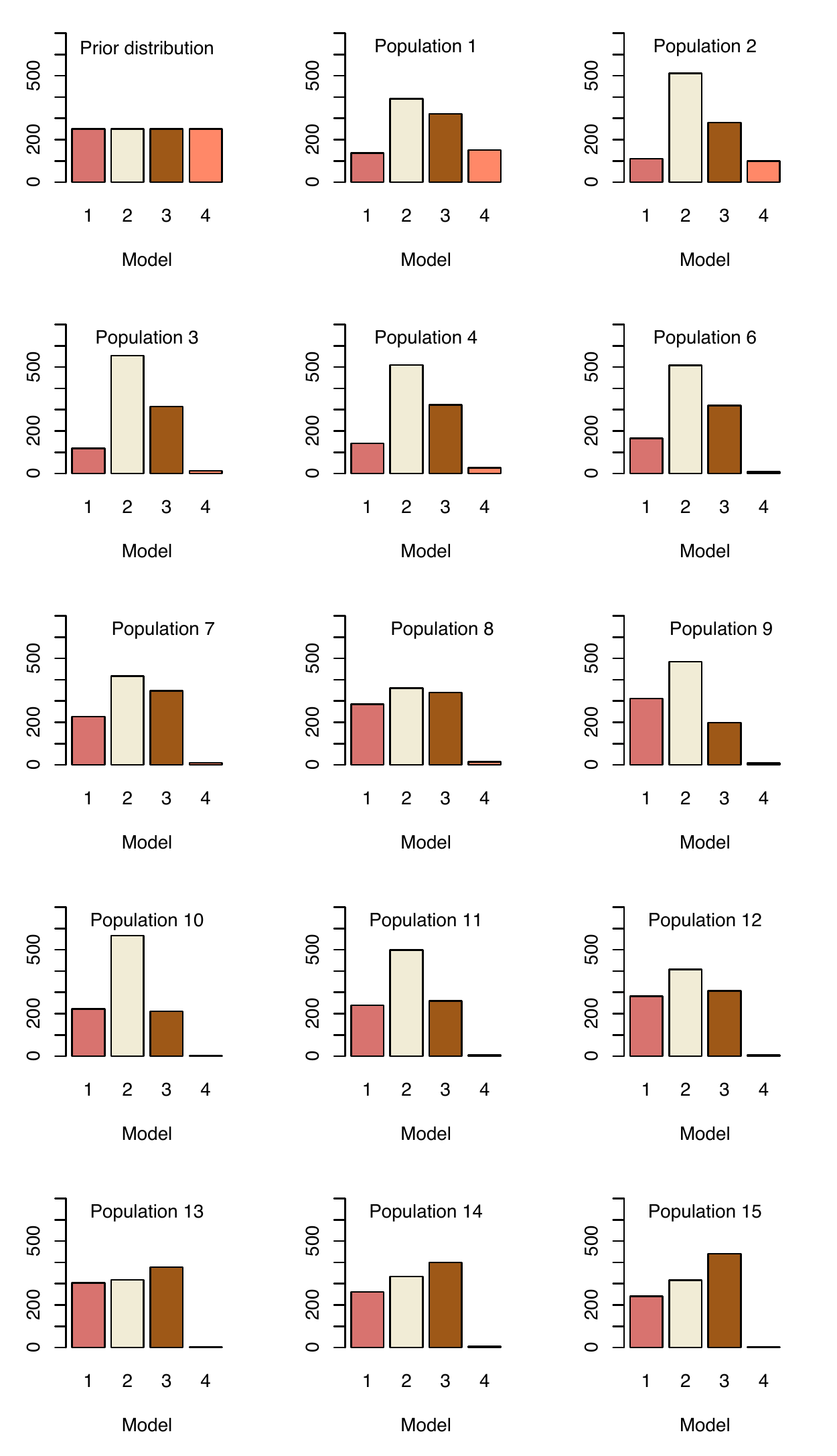}
	\end{center}
	\caption{\label{model_selection_real}\small{Populations of the marginal posterior distribution of $m$. Models $1$ to $4$ correspond to equations (\ref{basic}) - (\ref{thirdmodel}), respectively. An interesting phenomena is observed in populations $2$ to $12$, where model $2$ has the highest probability, in contrast to model $3$ having the highest inferred probability in the last population. The most probable explanation for this is that a local maximum favouring model $2$ has been passed on route to a global maximum of the posterior probability favouring model $3$.}}
\end{figure} 

\begin{figure}[h]	
	\begin{center}
	\includegraphics[width=70mm]{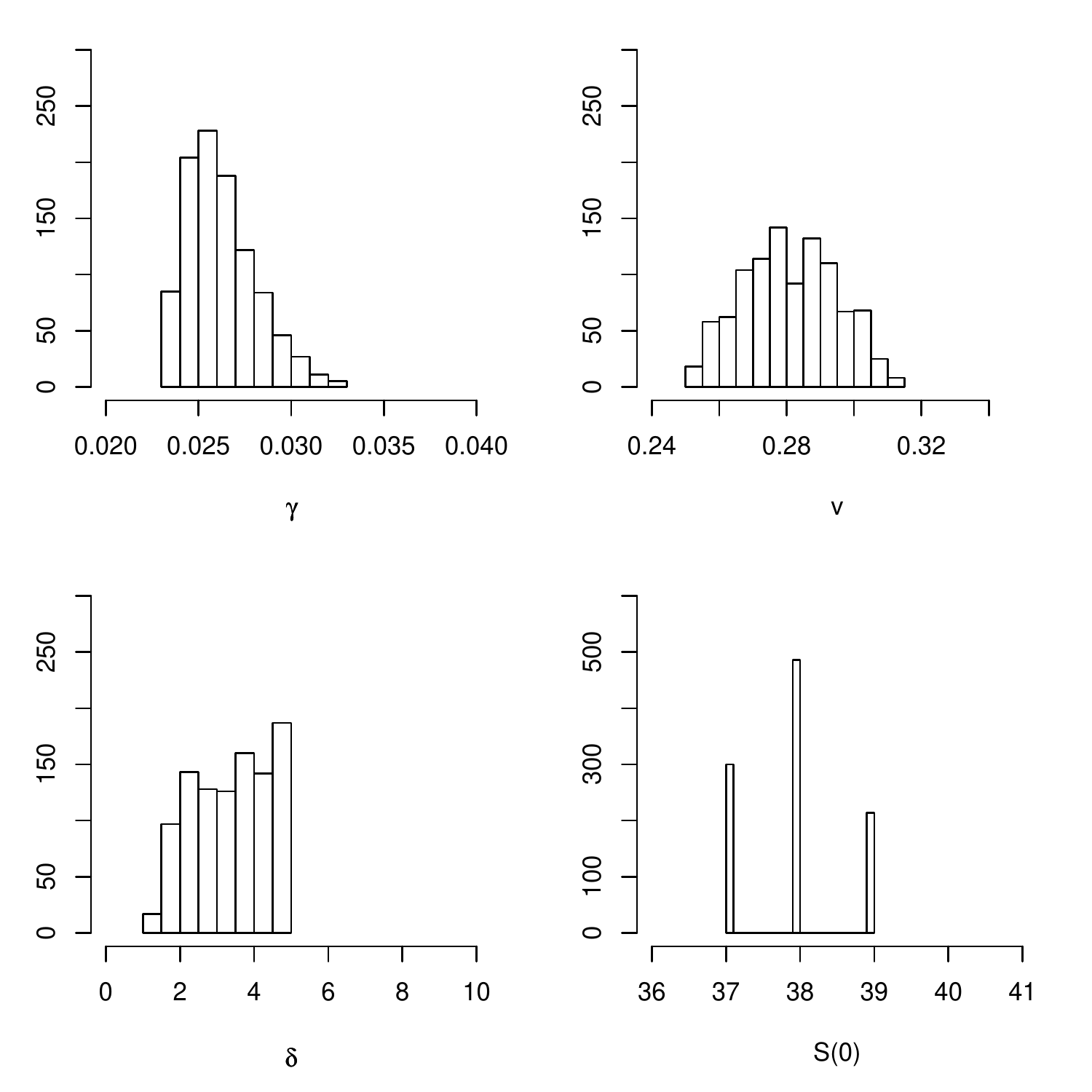}
	\end{center}
	\caption{\label{posteriors_SIR}\small{Histograms of the approximated posterior distributions of parameters $\gamma$, $v$, $\delta$ and $S(0)$  of the SIR model with the latent phase of infection (\ref{latent}).}}
\end{figure}

The target and intermediate distributions of model parameters are shown in Figure \ref{model_selection_real}.
The model selection algorithm chooses model (\ref{latent}), \ie the model with a latent class of disease carriers, to be the most suitable model for describing the data, however, it is only marginally better than models (\ref{basic}) and (\ref{delay}). Therefore, to draw reliable conclusions from the inferred parameters one might wish to use model averaging over models (\ref{basic}), (\ref{delay}) and (\ref{latent}). The marginal posterior distributions for parameters of model (\ref{latent}) are shown in Figure \ref{posteriors_SIR}.  However, as pointed out by \cite{Cook:2008p16606}, the estimated initial susceptible population size $S(0)$ is low compared to the whole population of the island, which suggests either that the majority of islanders were immune to a given strand of cold or (perhaps more plausibly) that the system is not well represented by our general epidemic models with homogeneous mixing. The estimated durations of the latent period, $\tau$ (from model (\ref{delay})) and $1/\delta$ (from model (\ref{latent})), however, are broadly in line with the established aetiology of the common cold \citep{NFields:1996p16691}. Thus within the set of candidate epidemiological models the ABC SMC approach selects the most plausible model and results in realistic parameter estimates.

\section{Discussion}

 Our study suggests that ABC SMC is a promising tool for reliable parameter inference and model selection for models of dynamical systems which can be efficiently simulated. Because of its simplicity and generality ABC SMC, unlike most other approaches, can be applied without change in deterministic and stochastic context (including models with time-delay).
\par
The advantage of Bayesian statistical inference, in contrast to most conventional optimization algorithms \citep{Moles:2003p8},  is that the whole probability distribution of the parameter can be obtained, rather than merely the point estimates of the optimal parameter values. 
Moreover in the context of hypothesis testing, the Bayesian perspective \citep{Cox:1974p15668,PRobert:2004p16815} has a more intuitive meaning than the corresponding frequentist point of view. ABC methods share these characteristics.
 \par
Another advantage of our ABC SMC approach is that observing the shape of intermediate and posterior distributions gives (without any further computational cost) information about the sensitivity of the model to different parameters and about the inferability of parameters. All  simulations are already part of the parameter estimation, and can be conveniently re-used for the sensitivity analysis via scatterplots or via the analysis of the posterior distribution using, \eg PCA. It can be concluded that the model is sensitive to parameters which are inferred quickly (in earlier populations) and which have narrow credible intervals, while it is less sensitive to the ones which get inferred in later populations and are not very localized by the posterior distribution. If the distribution does not change much between populations and resembles the uniform distribution from population 0, then it can be concluded that the corresponding parameter is not inferable given the available data.
\par
While parameter estimation for individual models is straightforward when a suitable number
of particles are used, more care should be taken in model selection problems; the domains of the uniform prior distributions should be chosen with care and acceptance rates should be closely monitored. These measures should prevent models being rejected in early populations solely due to inappropriately chosen (\eg too large) prior domains. Apart from potentially strong
dependence on the chosen prior distributions (which is also inherent in standard Bayesian model selection \citep{Kass:1995p2898}) we also observe dependency of Bayes factors to changes in the tolerance levels, $\epsilon_t$,  and perturbation kernel variances. Therefore care needs to be taken when applying the ABC SMC model selection algorithm.
\par
Finally, we want to stress the importance of monitoring convergence in ABC SMC. There are several ways to see if a good posterior distribution has been obtained: we can use inter-quartile ranges or tests of goodness-of-fit between successive intermediate distributions. A further crucial signature is the number of proposals required to obtain a specified number of accepted particles. This will also impose a practical limit on the procedure.
\par
For the problems in this paper the algorithm was efficient enough. Examples here were chosen to highlight different aspects of ABC SMC's performance and usability. However, for the use on larger systems the algorithm can be made more computationally
efficient by optimizing the number of populations, the distance function, the number of particles and perturbation kernels (\eg adaptive kernels). Moreover, the algorithm is easily parallelized. 

\section{Conclusion}

We have developed a sequential Monte Carlo ABC algorithm which can be used to estimate model parameters (including their credible intervals) and to distinguish among a set of competing models. This approach can be applied to deterministic and stochastic systems
in the physical, chemical and biological sciences, for example biochemical reaction networks or signaling networks. Because of the link between sensitivity and inferability ABC SMC can, however, also be applied to larger systems: critical parameters will be identified quickly, while the system is found to be relatively insensitive to parameters that are hard to infer. 


\section{Acknowledgments}
Financial support from the Division on Molecular Biosciences (Imperial College), MRC and the Slovenian Academy of Sciences and Arts (to T.T.), EPSRC (to D.W.), Wellcome Trust (to N.S., A.I. and M.P.H.S.) and EMBO (to M.P.H.S.) is gratefully acknowledged. We thank Ajay Jasra and James Sethna for helpful discussions and Paul Kirk, David Knowles and Sarah Langley for commenting on this manuscript. We are especially grateful to Mark Beaumont for providing helpful comments on an earlier version of the manuscript.

\appendix

\section{Appendix: Derivation of ABC SMC}

Here we derive the ABC SMC algorithm from the sequential importance sampling (SIS) algorithm of \cite{DelMoral:2006p4,DelMoral:2008p10336} and in Appendix B we show why this improves on the ABC partial rejection control (PRC) algorithm developed by \cite{Sisson:2007p2}. We start by briefly explaining the basics of importance sampling and present the SIS algorithm. 
\par
Let $\pi$ be the distribution we want to sample from, our \textit{target distribution}. If it is impossible to sample from $\pi$ directly, one can sample from a suitable \textit{proposal distribution}, $\eta$, and use importance sampling weights to approximate $\pi$. The Monte Carlo approximation of $\eta$ is 
$$ 
\hat{\eta}_N(x) = \frac{1}{N} \sum_{i=1}^{N}\delta_{X^{(i)}}(x),
$$
where $X^{(i)} \sim^{iid} \eta$ and $\delta_{x_0}(x)$ is the Dirac delta function, defined by
$$
\int_x f(x)\delta_{x_0}(x) dx = f(x_0).
$$
If we assume $\pi(x) > 0 \Rightarrow \eta(x)>0$, then the target distribution $\pi$ can be approximated by
$$
\hat{\pi}_N(x) = \frac{1}{N}\sum_{i=1}^{N} w(X^{(i)})\delta_{X^{(i)}}(x),
$$
with \textit{importance weights} defined as
$$
w(x) = \frac{\pi(x)}{\eta(x)}.
$$
In other words, to get a sample from the target distribution $\pi$,  one can instead sample from the proposal distribution, $\eta$, and weight the samples by importance weights, $w$.

In sequential importance sampling one reaches the target distribution $\pi_T$ through a series of intermediate distributions,  $\pi_t$, $t = 1,\ldots,T-1$. If it is hard to sample from these distributions one can use the idea of importance sampling described above to sample from a series of proposal distributions $\eta_t$ and weight the obtained samples by importance weights 
\begin{equation} \label{seq_weights} 
w_t(x_t) = \frac{\pi_t(x_t)}{\eta_t(x_t)}. 
\end{equation}
In SIS the proposal distributions are defined as
\begin{equation} \label{eq:proposal_distributions}
\eta_t(x_t) = 
\int \eta_{t-1}(x_{t-1}) \kappa_t(x_{t-1},x_t) dx_{t-1},
\end{equation}
where $ \eta_{t-1}$ is the previous proposal distribution and $\kappa_t$ is a Markov kernel. 
\par
%
In summary we can write the SIS algorithm as follows:

\begin{description}

 	\item [Initialization] 
	Set $t =1$.\\
	For $i = 1,\ldots, N$ draw $X_1^{(i)} \sim \eta_1$.\\
	Evaluate $w_1(X_1^{(i)})$ using (\ref{seq_weights}) and normalize.

	\item [Iteration] 
	Set $t = t + 1$, if $t = T+1$ stop.\\
	For $i = 1,\ldots,N$ draw $X_{t} \sim \kappa_t(X_{t-1}^{(i)},.)$.\\
	Evaluate $w_t(X_t^{(i)})$ using (\ref{seq_weights}) with $\eta_t(x_t)$ from (\ref{eq:proposal_distributions}) and normalize.\\
	
\end{description}

To apply SIS, one needs to define the intermediate and the proposal distributions. Taking this SIS framework as a base, we now define ABC SMC to be a special case of the SIS algorithm, where we choose the intermediate and proposal distributions in an ABC fashion, as follows. The intermediate distributions are defined as
$$
\pi_t(x) = \frac{\pi(x)}{B_t}\sum_{b=1}^{B_t}\mathbb{1} \left( d(D_0,D_{(b)}(x)) \leq \epsilon_t \right),
$$
where $\pi(x)$ denotes the prior distribution and $D_{(1)}, \ldots, D_{(B_t)}$ are $B_t$ datasets generated for a fixed parameter $x$,
$D_{(b)} \sim p(D|x)$. $\mathbb{1}(x)$ is an indicator function and  $\epsilon_t$ is the tolerance required from particles contributing to the intermediate distribution  $t$. This allows us to define $b_t(x) = \sum_{b=1}^{B_t}\mathbb{1} \left( d(D_0,D_{(b)}(x)) \leq \epsilon_t \right)$. We define the first proposal distribution to equal the prior distribution, $\eta_1 = \pi$. The proposal distribution at time $t$ ($t=2,\ldots,T$), $\eta_t$, is defined as the perturbed intermediate distribution at time $t-1$, $\pi_{t-1}$, such that for every sample, $x$, from this distribution we have \\
(i) $b_t(x)>0$ (in other words, particle $x$ is accepted at least one out of $B_t$ times), and \\
(ii) $\pi(x) >0$ (in order for a condition $\pi_t(x) > 0 \Rightarrow \eta_t(x)>0$ to be satisfied):
\begin{eqnarray} 
\eta_t(x_t) &=& \mathbb{1} \left( \pi_t (x_t)>0 \right) \mathbb{1} \left( b_t(x_t)>0 \right) \int \pi_{t-1}(x_{t-1}) K_t(x_{t-1},x_t)  dx_{t-1}  \label{prop_abc} \\
&=&   \mathbb{1} \left( \pi_t (x_t)>0 \right) \mathbb{1} \left( b_t(x_t)>0 \right) \int \eta_{t-1}(x_{t-1}) w_{t-1}(x_{t-1}) K_t(x_{t-1},x_t) dx_{t-1}, \nonumber
\end{eqnarray}
where  $K_t$ denotes the perturbation kernel (random walk around the particle). 

To calculate the weights defined by 
\begin{equation}
w_t(x) = \frac{\pi_t(x_t)}{\eta_t(x_t)} \label{eq:w},
\end{equation}
we need to find an appropriate way of evaluating $\eta_t(x_t)$ defined in equation (\ref{prop_abc}). This can be achieved through the standard Monte Carlo approximation, 
\begin{eqnarray}
\eta_t(x_t) &=&  \mathbb{1} \left( \pi_t (x_t)>0 \right) \mathbb{1} \left( b_t(x_t)>0 \right) \int \pi_{t-1}(x_{t-1}) K_t(x_{t-1},x_t)   d x_{t-1}  \nonumber  \\
&\approx&  \mathbb{1} \left( \pi_t (x_t)>0 \right) \mathbb{1} \left( b_t(x_t)>0 \right)  \frac{1}{N}\sum_{x_{t-1}^{(i)} \sim \pi_{t-1}} K_t(x_{t-1}^{(i)},x_t)  \nonumber,
\end{eqnarray}
where $N$ denotes the number of particles and $\{x_{t-1}^{(i)}\}$, $i = 1,\ldots, N$, are all the particles from the intermediate distribution $\pi_{t-1}$. The unnormalized weights
(\ref{eq:w}) can then be calculated as
\begin{equation} \nonumber
w_t(x_t) = \frac{\pi(x_t) b_t(x_t) }{ \sum_{x_{t-1}^{(i)} \sim \pi_{t-1}} K_t(x_{t-1}^{(i)},x_t) },
\end{equation}
where $\pi$ is the prior distribution, $b_t(x_t) = \sum_{b=1}^{B_t} \mathbb{1} \left( d(D_0,D_{(b)}(x_t)) \leq \epsilon_t \right)$, and $D_{(b)}$, $b=1,\ldots,B_t$, are the $B_t$ simulated  datasets for a given parameter $x_t$. For $B_t =1$ the weights become
\begin{equation} \nonumber
w_t(x_t) = \frac{\pi(x_t)}{\sum_{x_{t-1}^{(i)} \sim \pi_{t-1}} K_t(x_{t-1}^{(i)},x_t)}
\end{equation}
for all \textit{accepted} particles $x_t$.

The ABC SMC algorithm can be written as follows: 
\begin{description}

 	\item [S1] Initialize $\epsilon_1, \ldots, \epsilon_T$.\\
Set the population indicator $t = 0$. 

	\item [S2.0] Set the particle indicator $i = 1$.

	\item [S2.1] If $t = 0$, sample $x^{**}$ independently from $\pi(x)$.\\
	If $t > 0$, sample $x^*$ from the previous population $\{x_{t-1}^{(i)}\}$ with weights $w_{t-1}$ and perturb the particle to obtain $ x^{**} \sim K_t(x|x^{*})$, where $K_t$ is a perturbation kernel.\\ 
	If $\pi(x^{**}) = 0$, return to \textbf{S2.1}.\\
	Simulate a candidate dataset $D_{(b)}(x^{**}) \sim f(D|x^{**})$ $B_t$ times ($b=1,\ldots,B_t$) and calculate $b_t(x^{**})$.\\
	If $b_t(x^{**})=0$, return to \textbf{S2.1}.
	
	\item [S2.2] Set $x_t^{(i)} = x^{**}$ and calculate the weight for particle $x_t^{(i)}$,
	$$
	w_t^{(i)} =
	\left\{ \begin{array}{ll}
	b_t(x_t^{(i)}),  &  \textrm{if } t=0,  \\
	\frac{\pi(x_t^{(i)})b_t(x_t^{(i)})}{ \sum_{j=1}^{N} w_{t-1}^{(j)} K_t(x_{t-1}^{(j)},x_t^{(i)})}, & \textrm{if } t >0.
	\end{array} \right.
	$$
	If $i < N$ set $i = i+1$, go to \textbf{S2.1}.

	\item [S3] Normalize the weights.\\
	If $t<T$, set $t = t+1$, go to \textbf{S2.0}.   

\end{description}
When applying ABC SMC to deterministic systems we take $B_t = 1$, \ie we simulate the dataset for each particle only once. 

\section{Appendix: Comparison of ABC SMC algorithm with the ABC PRC of Sisson \textit{et al.}}

In this section we contrast the ABC SMC algorithm, which we developed above, and the ABC PRC algorithm of  \cite{Sisson:2007p2}. The algorithms are very similar in principle, and the main difference is that we base ABC SMC on  a SIS framework whereas Sisson \textit{et al.} use a SMC sampler as a basis for ABC PRC, where the weight calculation is done through the use of a backward kernel. Both algorithms are explained in detail in \cite{DelMoral:2006p4, DelMoral:2008p10336}. The disadvantage of the SMC sampler is that it is impossible to use an optimal backward kernel and it is hard to choose a good one. Sisson \textit{et al.} choose a backward kernel that is equal to the forward kernel, which we suggest can be a poor choice\footnote{See also http://web.maths.unsw.edu.au/\~{}scott/papers/paper\_smcabc\_optimal.pdf.}. While this highly simplifies the algorithm since in the case of a uniform prior distribution all the weights become equal, the resulting posterior distributions can result in bad approximations to the true posterior. In particular, using equal weights can profoundly affect the posterior credible intervals\footnote{This, including the experiment below, was suggested by \cite{Beaumont:2008p17400}.}.

\begin{figure}[th]	
	\centering
	\includegraphics[width = 160mm]{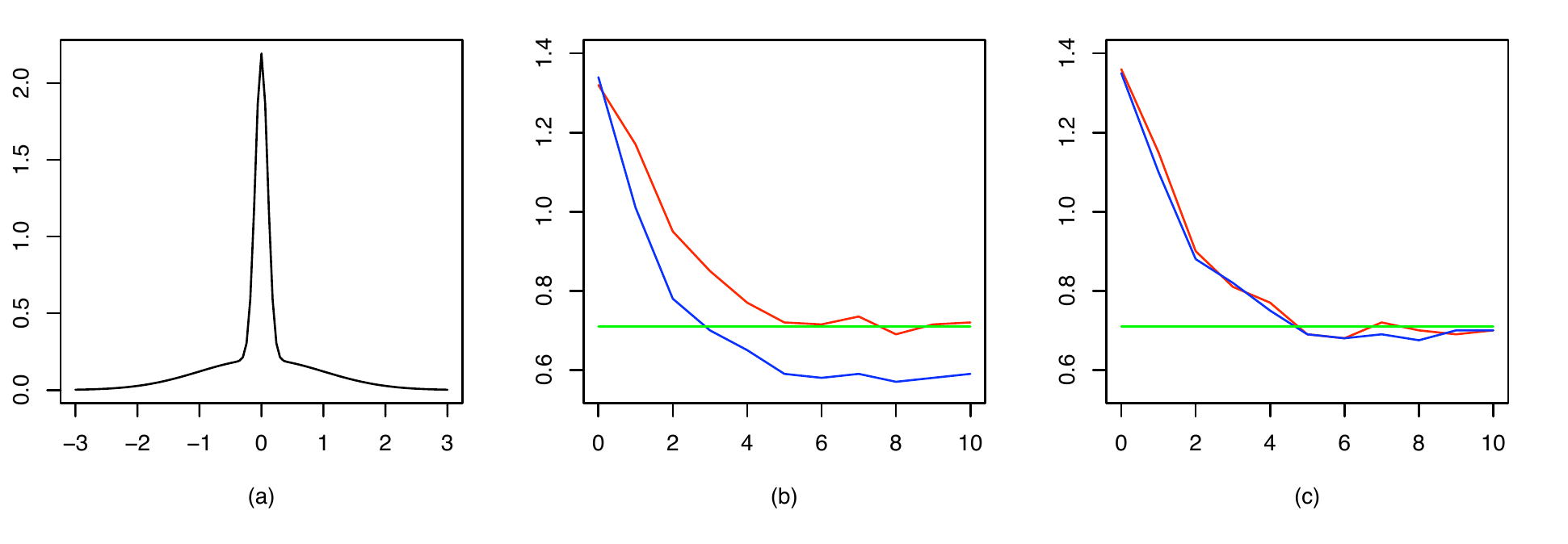}
	\caption{\small{(a) The probability density function of a mixture of two normal distributions, $\frac{1}{2}\phi \left( 0,\frac{1}{100} \right) + \frac{1}{2}\phi \left(0,1 \right)$, taken as a toy example used for comparison of ABC SMC with ABC PRC. (b)-(c) Plots show how the variance of approximated intermediate distributions $\pi_t$ changes with populations ($t=1,\ldots,10$ on x-axis). The red curves plot the variance of ABC SMC population and the blue curves the variance of non-weighted ABC PRC populations. The perturbation kernel in both algorithms is uniform, $K_t = \sigma U(-1,1)$. In (b) $\sigma=1.5$ and this results in poor estimation of the posterior variance with ABC PRC algorithm. In (c) $\sigma$ is updated in each population so that it expands over the whole population range. Such $\sigma$ is big enough for non-weighted ABC PRC to perform equally well as ABC SMC.} \label{new_std_average}}
\end{figure}

Here we compare the outputs of both algorithms using the toy example from \cite{Sisson:2007p2}. The goal is to generate samples from a mixture of two normal distributions,
$
\frac{1}{2}\phi \left( 0,\frac{1}{100} \right) + \frac{1}{2}\phi \left(0,1 \right),
$
where $\phi(\mu,\sigma^2)$ is the density function of a $N(\mu,\sigma^2)$ random variable (Figure \ref{new_std_average}(a)). Sisson \textit{et al.} approximate the distribution well by using three populations with $\epsilon_1 = 2$, $\epsilon_2 = 0.5$ and $\epsilon_3 = 0.025$, respectively, starting from a uniform prior distribution. However, ABC PRC would perform poorly if they had used more populations. In Figures \ref{new_std_average}(b) and \ref{new_std_average}(c) we show how the variance of the approximated posterior distribution changes through populations. We use $100$ particles and average over $30$ runs of the algorithm, with tolerance schedule $\epsilon = \{2.0, 1.5, 1.0, 0.75, 0.5, 0.2, 0.1, 0.075, 0.05, 0.03, 0.025 \}$. The red line shows the variance from ABC SMC, the blue line the variance from the ABC PRC, and the green line the variance from the ABC rejection algorithm. In case when the perturbation is relatively small, we see that the variance resulting from improperly weighted particles in ABC PRC is too small (blue line in Figure \ref{new_std_average}(b)), while the variance resulting from ABC SMC is ultimately comparable to the variance obtained by ABC rejection algorithm (green line). 

However, we also notice that using many populations and a too small perturbation, \eg uniform perturbation $K_t = \sigma U(-1,1)$ with $\sigma=0.15$, the approximation is not very good (results not shown). One therefore needs to be careful to use a sufficiently large perturbation, irrespective of the weighting scheme. However, at the moment there are no strict guidelines of how best to do this and we decide on the perturbation kernel based on experience and experimentation.

For the one-dimensional deterministic case with a uniform prior distribution and uniform perturbation, one can show that all the weights will be equal when $\sigma$ of the uniform kernel is equal to the whole parameter range. In this case the non-weighted ABC PRC yields the same results as ABC SMC  (Figure \ref{new_std_average}(c)). However, when working with complex high-dimensional systems it simply is not feasible to work with only a small number of populations, or a very big perturbation kernel spreading over the whole parameter range. Therefore we conclude that it is important to use the ABC SMC algorithm because of its correct weighting. 

Another difference between the two algorithms is that ABC PRC includes the resampling step in order to maintain a sufficiently large effective sample size (ESS, \citep{DelMoral:2006p4, Liu:1998p17086, Liu:2001p16897}). In contrast to non-ABC SIS or SMC frameworks, ABC algorithms, by sampling with weights in step \textbf{S2.1} prior to perturbation, we suggest, do not require this additional resampling step.  

%
We note that in SIS the evaluation of weights after each population is computationally more costly, \ie $O(N^2)$, than the calculation of weights in SMC with a backward kernel, which is  $O(N)$, where $N$ denotes the number of particles. However, this cost is negligible in the ABC case, because the vast proportion of  computational time in ABC is spent on simulating the model repeatedly. While thousands or millions of simulations are needed, the weights only need to be updated after every population. Therefore we can easily afford spending a bit more computational time in order to use the correctly weighted version and circumvent the issues related to backward kernel choice.

\section{Appendix: ABC and full likelihood for ODE systems}

We would like to note that the approximate Bayesian computation algorithms with distance function chosen to be squared errors is equivalent to
the maximum likelihood problem for a dynamical systems for which Gaussian errors are assumed. In other words, minimizing the distance function
\begin{equation} 
\sum_{i=1}^{n}\sum_{j=1}^{m} \left(  x_{ij}- g_j(t_i,\theta)\right)^2 ,
\end{equation}
where $g(t,\theta) \in  {\mathbb R}^{m}$ is the solution of the $m$-dimensional dynamical system and
$D = (x_{i})_{\{i=1,\ldots,n\}}$, $x_{i} \in  {\mathbb R}^{m}$, are (m-dimensional) data points measured at times $t_1,\ldots,t_n$,
is equivalent to maximizing the likelihood function
$$
\prod_{i=1}^{n} \frac{1}{(2\pi)^{m/2}|\Sigma|^{\frac{1}{2}}} e^ { -\frac{1}{2}(x_i- g(t_i,\theta))^T \Sigma^{-1} (x_i - g(t_i,\theta)) },
$$
where $\Sigma$ is diagonal and its entries equal. This can straightforwardly be generalized for the case of multiple time series measurements. Thus ABC is for deterministic ODE systems closely related to standard Bayesian inference where the likelihood is evaluated. This is because ODEs are not based on a probability model, and likelihoods are therefore generally defined in a non-linear regression context (such as assuming that data is normally distributed around the deterministic solution), see \eg \cite{Timmer:2004p474} and \cite{Vyshemirsky:2008p14865}. 

\section*{Supplementary material}

\subsection*{Video of prior, intermediate and posterior distributions of deterministic repressilator model}

The video shows the particles from the $4$-dimensional distributions corresponding to the (a) prior distribution, (b) $4^{th}$ intermediate distribution, (c) approximate posterior distribution obtained for the deterministic repressilator model. The video was produced using the GGobi \citep{Swayne:2003p15443} and RGGobi \citep{TempleLang:2006p15507} packages from the R statistical environment (\url{www.R-project.org}). 


\subsection*{Datasets}
Simulated datasets used in examples in this paper are attached as a supplementary material.



\end{document}